\documentclass[letterpaper]{article}

\usepackage{emulateapj}
\usepackage{onecolfloat}
\usepackage{apjfonts}
\usepackage{amsmath}
\usepackage{graphicx}

\begin{document}

\submitted{To Appear in the Astrophysical Journal}

\twocolumn[
\title{The X-ray Properties of Low-Frequency Quasi-Periodic 
Oscillations from GRS 1915+105 up to 120\,keV}

\authoremail{jtomsick@ucsd.edu}

\author{John A. Tomsick}
\affil{Center for Astrophysics and Space 
Sciences, University of California, San Diego, MS 0424, La Jolla, CA 92093 
(e-mail: jtomsick@ucsd.edu)}

\vspace{0.1cm}

\author{Philip Kaaret}
\affil{Harvard-Smithsonian Center for Astrophysics, 60 Garden Street, 
Cambridge, MA 02138 (e-mail: pkaaret@cfa.harvard.edu)}

\begin{abstract}

We present a study of the properties of strong 0.8-3.0\,Hz 
quasi-periodic oscillations (QPOs) that occurred during
1997 {\em Rossi X-ray Timing Explorer} ({\em RXTE})
observations of the microquasar GRS~1915+105 in the low-hard 
state.  The high count rates allow us to track individual 
QPO peaks, and we exploit this to develop a QPO folding 
technique.  In contrast to previous QPO studies with 
{\em RXTE}, we emphasize the high energy QPO properties and 
report the detection of a QPO in the 60-124\,keV energy band.  
Our technique allows us, for the first time, to measure the 
phase of the QPO harmonics relative to the fundamental.  
Variation in this phase difference leads to changes in the 
shape of the QPO profile with energy and over time.  The 
strength of the QPO fundamental increases up to 19\,keV,
but the data do not suggest that the strength continues
to increase above this energy.  In some cases, the QPO
amplitudes in the 30-60\,keV and 60-124\,keV energy bands
are significantly less than in the 13-19\,keV and 19-29\,keV 
energy bands.  We also use our technique to measure the phase 
lag of the QPO fundamental and harmonics.  In the case where 
negative phase lags are detected for the fundamental, positive 
phase lags are detected for the first harmonic.

\end{abstract}


] 

\section{Introduction}

The {\em Rossi X-ray Timing Explorer} (\cite{brs93}) has provided 
an unprecedented opportunity to study the timing properties of 
X-ray binaries.  {\em RXTE} observations of black hole candidates 
(BHCs) indicate that modulations in the X-ray flux 
(i.e., quasi-periodic oscillations or QPOs) occur on time scales from 
milliseconds to hundreds of seconds.  In this paper, we focus on 
QPOs from one of the most studied {\em RXTE} targets, GRS~1915+105.  
This BHC X-ray transient and microquasar (\cite{mr94}) has the 
richest variety of timing behavior of any X-ray binary observed to 
date (e.g., \cite{gmr96};~\cite{belloni2000}).  

Previous studies of GRS~1915+105 show that the QPO properties are
closely related to the properties of the energy spectrum.  In
these studies, the energy spectrum was modeled using a two component
model consisting of a soft component, which presumably comes from
an optically thick accretion disk, and a hard component, which is
thought to be due to inverse Comptonization of soft photons from
the accretion disk.  Markwardt et al.~(1999) and Muno et al.~(1999) 
find that the QPO frequency is closely related to the properties of 
the soft component and, thus, the accretion disk.  It has also been
found that the strength of the QPO increases with energy up to at 
least 15\,keV, indicating that the QPO mechanism modulates the hard 
component (e.g., \cite{mrg97}).  If the hard component is produced 
via inverse Comptonization, a hard phase lag is expected.  While 
hard phase lags have been observed for several sources, 
Reig et al.~(2000) find soft phase lags for GRS~1915+105 in some 
cases, and, in this paper, we include some of the observations
where soft phase lags were observed.

We present a detailed study of 0.8-3.0\,Hz QPOs that occurred during
1997 October {\em RXTE} observations.  The high Proportional Counter
Array (PCA) count rates allow us to track individual QPO peaks, and 
we have exploited this to develop a QPO folding technique with 
some similarities to that used by Morgan et al.~(1997).  In contrast 
to previous QPO studies with {\em RXTE}, we emphasize the high energy 
QPO properties and report the detection of a QPO in the 60-124\,keV
band using the High Energy X-ray Timing Experiment (HEXTE).  Here, 
we study how the properties of the QPO fundamental and first and
second harmonics change with energy and between observations.  We 
measure the phase of the QPO harmonics relative to the fundamental 
and the strength and phase of the fundamental and harmonics over the 
full {\em RXTE} band-pass.

\begin{figure}[t]
\plotone{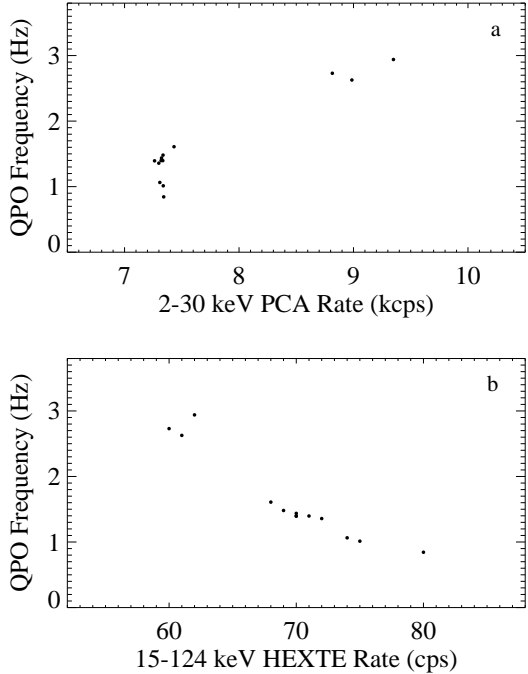}
\vspace{0.5cm}
\caption{QPO frequency vs.~count rate for (a) the PCA in
the 2-30\,keV band and (b) HEXTE in the 15-124\,keV band.
The numbers used to generate these plots are given in
Table~\ref{tab:obs}.\label{fig:f_vs_r}}
\end{figure}

\section{Observations}

We chose the 1997 October {\em RXTE} observations from the large 
number of GRS~1915+105 observations between the start of the {\em RXTE} 
mission and early 1998 based on our goal of extending QPO measurements 
to high energies.  We searched the public archive for observations with 
strong low-frequency QPOs and a high HEXTE count rate, and found that 
observations during low-hard states (Morgan et al.~1997) provide the best 
match to these criteria.  Between the start of the {\em RXTE} mission 
and early 1998 the low-hard state occurred three times for GRS~1915+105 
as shown in Figure~9 of Belloni et al.~(2000) along with the {\em RXTE} 
All-Sky Monitor (ASM) light curve\footnote{%
Note that Belloni et al.~(2000) refer to the low-hard state as 
class $\chi$.}  In the 1.5-12 keV ASM light curve, the low-hard
states appear as low flux time periods with very little variability.
Ultimately, we chose to focus on the third low-hard state.  This
decision was made because HEXTE was not rocking during the first
low-hard state, making background subtraction with standard
techniques impossible, and because the QPO frequency is lower for 
the third low-hard state than for the second low-hard state 
(\cite{tcg98};~\cite{reig2000}).  Since our folding technique 
depends on being able to detect individual QPO peaks, low frequency 
QPOs provide a significant advantage.

During the third low-hard state, six {\em RXTE} observations were 
made over thirteen {\em RXTE} orbits.  For each of the thirteen 
orbits, Table~\ref{tab:obs} gives the time at the midpoint of 
the unocculted part of the orbit, the exposure time, the QPO 
frequency during the orbit, which we determined by making and 
fitting power spectra for each orbit, the 2-30\,keV PCA count 
rate and the 15-124\,keV HEXTE count rate.  The PCA count rates 
are between 7300 and 9500 cps, and the HEXTE count rates are 
between 60 and 80 cps.  Figure~\ref{fig:f_vs_r} shows the QPO 
frequencies as a function of count rate for PCA and HEXTE.  The 
correlation between the PCA rate and the QPO frequency is apparent 
and has been noticed before for this data (\cite{reig2000}).  
Interestingly, there is an anti-correlation between the HEXTE 
rate and the QPO frequency that is even more striking.  The 
2-30\,keV PCA light curve for 15\,s of orbit 4 is shown in 
Figure~\ref{fig:lightcurve}a, and the presence of the QPO is 
obvious.

\section{Analysis}

\begin{figure}
\plotone{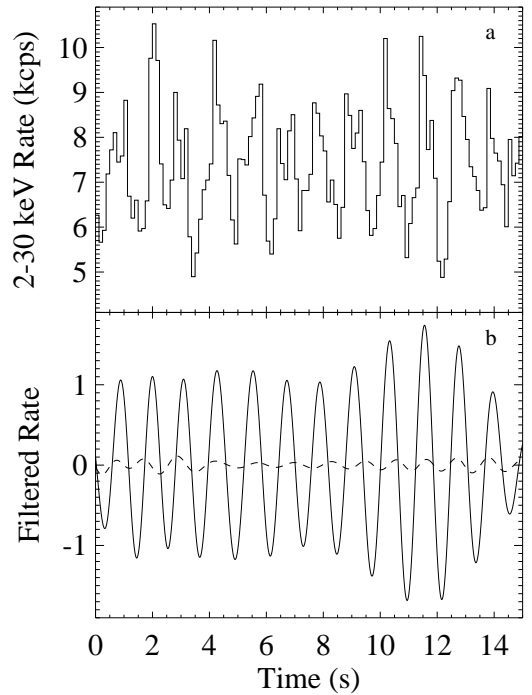}
\vspace{0.5cm}
\caption{(a) The 2-30\,keV PCA light curve for 15\,s of orbit 4
with a time resolution of 0.125\,s.  (b) The same light curve after 
band-pass filtering (solid line).  For comparison, the dashed line
shows a white noise light curve after band-pass filtering.
\label{fig:lightcurve}}
\end{figure}

For each {\em RXTE} orbit, we extracted the 2-30\,keV PCA light 
curve with 7.8125\,ms time resolution.  We made and fitted power 
spectra to determine QPO frequencies (see Table~\ref{tab:obs}) 
for each orbit.  A band-pass filter with frequency limits 20\% 
below and above the QPO frequency was applied to the light curve, 
and the filtered light curve for 15\,s of orbit 4 is shown in 
Figure~\ref{fig:lightcurve}b.  We used a Kaiser band-pass filter, 
which is a finite impulse response filter with a Kaiser window 
(\cite{jackson89}).  Derivatives of the filtered light curve 
were calculated to find the minima between QPO pulses.  To 
determine which minima are significant, we simulated a light curve 
with a white noise power spectrum and the same statistics as the 
observed light curve.  We applied the band-pass filter to the 
white noise light curve, and the result is the dashed line in 
Figure~\ref{fig:lightcurve}b.  In the filtered version of the 
observed light curve, minima are defined as the places where the 
first derivative of the filtered light curve is zero, the second 
derivative of the filtered light curve is positive and the minimum 
is lower than 99.9\% of the minima in the filtered white noise
light curve.  

\begin{figure}[t]
\plotone{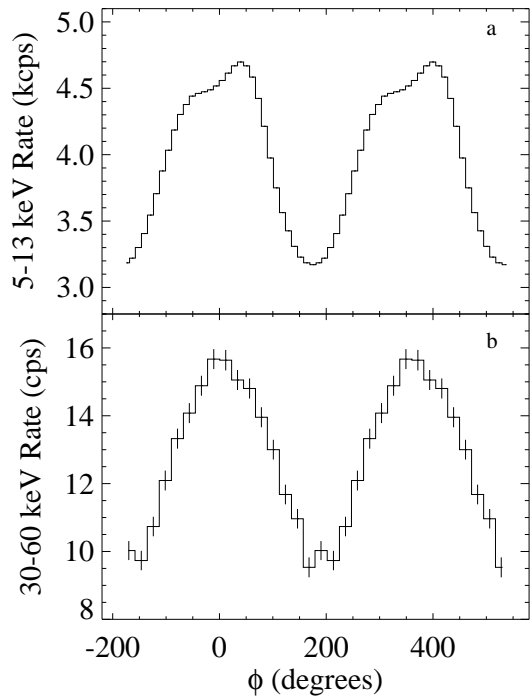}
\vspace{0.5cm}
\caption{Folded (a) PCA and (b) HEXTE light curves for the 
combination of orbits 7-13.  The 5-13\,keV PCA and the 30-60\,keV 
HEXTE energy bands are shown, and two QPO cycles are displayed.
\label{fig:folded}}
\end{figure}

The minima from the 2-30\,keV PCA light curve are used to fold the 
light curves in different energy bands.  We extracted PCA and 
HEXTE light curves in several energy bands.  The standard deadtime 
correction is applied to the HEXTE data.  To fold the light curves, 
we assign a pulse phase to each light curve time bin.  We define the 
light curve minima to be at a pulse phase of $180^{\circ}$ and 
determine phases for the other time bins by interpolating between 
minima.  A plot of count rate vs.~phase provides the folded light 
curve.  This method allows us to combine data from orbits with 
slightly different QPO frequencies.  Representative folded light 
curves for PCA and HEXTE are shown in Figure~\ref{fig:folded}.  
After folding as described above, the light curve is rebinned in 
phase.  Poisson error bars are shown and two cycles are displayed.

To optimize and test our software, we applied our folding technique 
to simulated light curves.  We used the GRS~1915+105 power
spectra to determine the properties of the simulated light curves.
We made and fitted power spectra for each of the thirteen orbits, 
and we refer the reader to Reig et al.~(2000) for figures showing 
a representative sample of these.  The power spectra contain a 
low-frequency (0.8-3.0\,Hz) QPO with a fractional rms amplitude 
of about 10\% and its first harmonic with an rms amplitude near 5\%.  
Our simulated light curves contain a sinusoidal modulation with
a frequency of 1\,Hz and an amplitude of 14\% of the mean level, 
which corresponds to an rms amplitude of 10\%.  A harmonic is 
included by adding a second sinusoid with twice the frequency and
half the strength of the first sinusoid.  A fixed but arbitrary phase
difference between the two sinusoids is assumed.  The signal is made 
to be quasi-periodic by stretching or compressing each cycle 
(i.e., pulse) in time.  Although the pulse profile remains constant 
throughout, the pulse duration changes for each cycle.  Motivated by 
the results of Morgan et al.~(1997), who found that the phase of the 
QPO relative to a periodic signal performs a random walk over time, 
we vary the pulse duration so that it performs a random walk from 
pulse-to-pulse.  Since a changing pulse duration leads to a 
corresponding change in the phase of the QPO, we set the step 
size for the random walk to reproduce the magnitude of the phase 
changes observed for GRS~1915+105 by Morgan et al.~(1997).  
Finally, we checked that our simulated light curves have properties 
consistent with those of the actual GRS~1915+105 light curves by 
making power spectra for the simulated light curves.  The power 
spectra contain a QPO and its harmonic, and the widths of these 
features are consistent with those found for the actual 
GRS~1915+105 power spectra.

\begin{figure}
\plotone{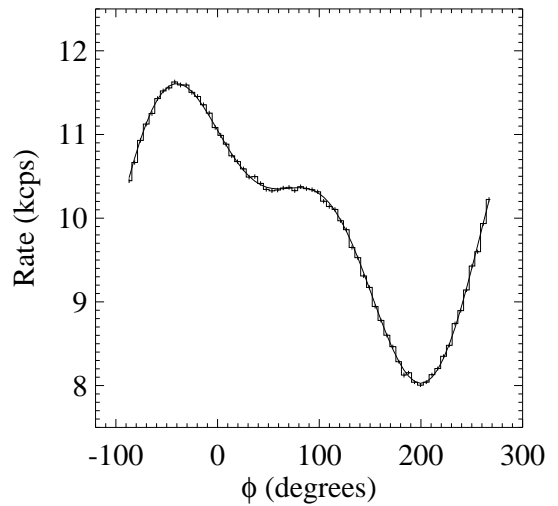}
\vspace{0.5cm}
\caption{Comparison between a folded light curve and the input QPO 
profile for the simulation.  The largest discrepancy between the 
folded light curve and the input profile for any phase bin is 
0.74\% of the mean count rate.\label{fig:sim}}
\end{figure}

We applied our QPO folding technique, which is outlined above, 
to the simulated light curves.  Figure~\ref{fig:sim} shows the 
result for a representative simulated light curve.  In this example, 
the largest discrepancy between the folded light curve and the 
input profile for any phase bin is 0.7\% of the mean count rate.  
For optimization of our technique, we filtered the simulated light 
curves using several different frequency limits for the band-pass.  
As long as the band-pass is wide enough to contain the 
random-walking QPO frequency, the results do not depend sensitively 
on the exact frequency limits used for the filter, and
frequency limits 20\% below and above the QPO frequency 
provided a band-pass wide enough for GRS~1915+105.  
We applied a band-pass filter with these frequency limits to
100 simulated light curves, and, for each light curve, we 
determined the largest discrepancy between the folded light curve 
and the input profile.  On average, the largest discrepancy was 
0.74\% of the mean count rate, and, for 95 of the 100 simulated 
light curves, the largest discrepancies were less than 1\%.  We 
produced simulated light curves with various phase differences 
between the QPO and its harmonic.  Although the shape of the 
QPO profile changes, the discrepancies between the folded light 
curve and the input profile do not change significantly.

Using our technique for finding the minima in the GRS~1915+105
light curve, 95\% to 99\% of the expected minima are detected.
This either indicates that there are irregularities in the 
light curve (e.g., the QPO turns off for a small fraction of
the time) or that our technique is not sensitive enough to 
detect a small fraction of the minima.  In order to determine 
where minima are being missed, we calculated the ratio of the 
time to the previous minimum to the time to the next minimum 
for each detected minimum.  The ratio is close to 1.0 in cases 
where there are no missing minima, but significantly deviates 
from 1.0 where missing minima occur.  To avoid including data
with indeterminate phases, we rejected data surrounding missing
minima.  In the worst case, this technique led to rejecting 8\% 
of the data, and, for most orbits, between 2\% and 4\% of the 
data were rejected.  

\section{Results}

\begin{figure}[t]
\plotone{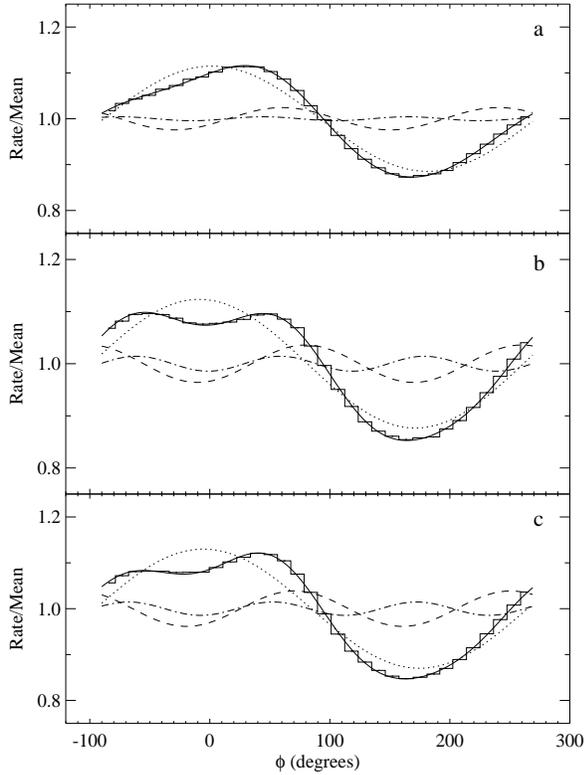}
\vspace{0.5cm}
\caption{The 2-5\,keV folded light curves (Rate/Mean) vs.~phase
for orbits 1-3 (a), 4-6 (b) and 7-13 (c).  The solid line is a
fit to the light curve with a model consisting of a constant
and three harmonically spaced sinusoids (see Equation 1).  The
fundamental (dotted line), first harmonic (dashed line) and
second harmonic (dot-dashed line) are also shown separately.
\label{fig:fits}}
\end{figure}

We produced folded light curves in five PCA energy bands (2-5\,keV, 
5-13\,keV, 13-19\,keV, 19-29\,keV and 29-60\,keV) and four HEXTE 
energy bands (16-18\,keV, 18-30\,keV, 30-60\,keV and 60-124\,keV) 
for each of the thirteen {\em RXTE} orbits.  We examined the 
folded light curves and found that the shape of the folded light 
curve changes during the thirteen orbits.  However, in some cases, 
the shape is very similar for consecutive orbits, and, in these 
cases, we combined the data from multiple orbits.  The longest stretch 
of time during which the shape remains stable is more than 3\,d during 
orbits 7-13.  Also, the QPO profile can change significantly on time 
scales less than 5 or 6\,d.  We characterize the folded light curves by 
fitting them with harmonically related sinusoids.  From 
Figure~\ref{fig:folded}, it is clear that one sinusoid does not provide 
a good description of the folded light curves, especially at low energy.  
Adding more sinusoids improves the fits considerably, and three 
sinusoids provide a good description of the data.  We note that the 
presence of a QPO with two harmonics is consistent with the power 
spectra shown by Reig et al.~(2000).  We fitted the folded light curves 
with the function
\begin{equation}
f(\phi) = a_{0} + a_{1}\sin(\phi-\phi_{1}) + a_{2}\sin[2(\phi-\phi_{2})] +
	a_{3}\sin[3(\phi-\phi_{3})]~~,
\end{equation}
and, below, we refer to $\phi_{i}$ as the phase of the sinusoid, 
and we define the amplitudes of the sinusoids as 
$A_{i} = a_{i}/a_{0}$, where $i$ runs from 1 to 4.  For most
energy bands, the phase-averaged deviations between this model 
and the folded light curve are less than 1\%.  The deviations are
somewhat larger in a few cases where the folded light curves have
relatively low statistics.  In these cases, low reduced chi-squared 
values ($\chi^{2}/\nu < 0.8$ for 25 degrees of freedom for the 
PCA and $\chi^{2}/\nu < 1.3$ for 9 degrees of freedom for 
HEXTE) indicate that the model provides a good description
of the data.

\begin{figure}
\plotone{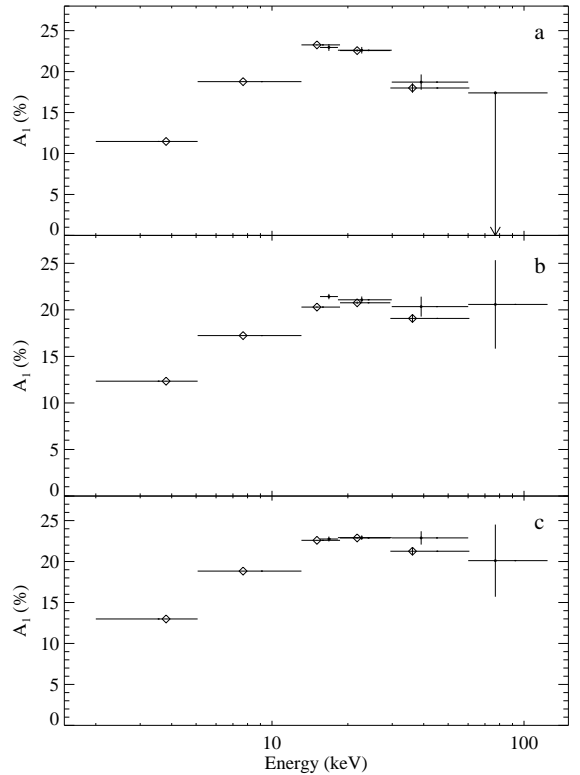}
\vspace{0.5cm}
\caption{The QPO amplitude of the fundamental ($A_{1}$) vs.~energy
for orbits 1-3 (a), 4-6 (b) and 7-13 (c).  The diamonds mark PCA 
measurements and the points mark HEXTE measurements.  The error bars 
shown correspond to 68\% confidence, and the 95\% confidence upper 
limit is shown for the 60-124\,keV energy band in panel a.
\label{fig:amp1}}
\end{figure}

In Figure~\ref{fig:fits}, we show the 2-5\,keV folded light curves 
for orbits 1-3, 4-6 and 7-13, fitted with the model described 
above.  The solid line shows the combination of three sinusoids,
and the fundamental (dotted line), first harmonic (dashed line)
and second harmonic (dot-dashed line) are also shown.  While 
the commonly used power spectrum does not provide information 
about the phase difference between sinusoidal components, our 
folding technique does provide this information.  Phase information 
can be derived from examination of the full complex Fourier 
transform, but this is only rarely accomplished and has not been 
done in analyses of GRS~1915+105.  Table~\ref{tab:phases} 
gives the values of $\phi_{1}-\phi_{2}$, $\phi_{1}-\phi_{3}$ 
and $\phi_{2}-\phi_{3}$ for the 2-5\,keV folded light curves.  
The shape of the folded light curve depends on the phase difference 
between the sinusoidal components.  The largest change in the 
shape of the folded light curve occurs because of changes in the 
phase difference between the fundamental and the two harmonics.  
The variation in $\phi_{2}-\phi_{3}$ over time is smaller.  In 
addition, we note that the double peaking observed for orbits 4-6 
occurs because the minima of the harmonics occur at nearly the 
same phase as the maximum of the fundamental.

\begin{figure}[t]
\plotone{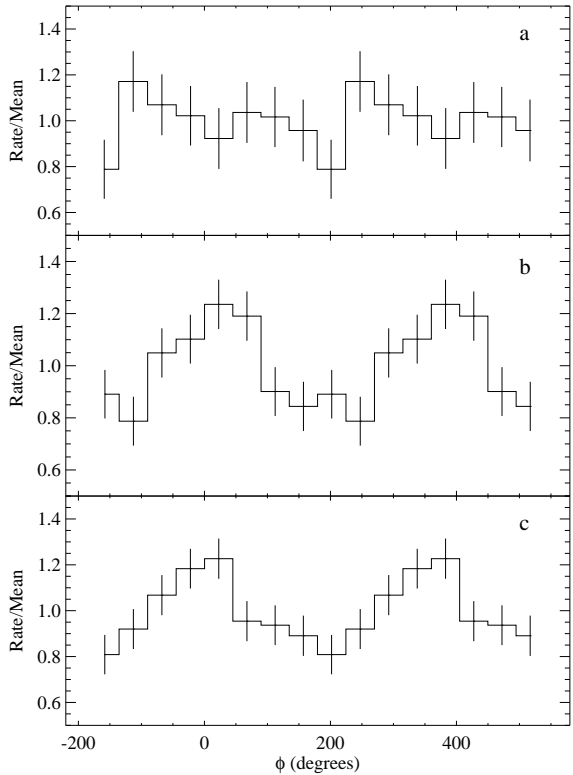}
\vspace{0.5cm}
\caption{The 60-124\,keV folded light curves (Rate/Mean) vs.~phase 
for orbits 1-3 (a), 4-6 (b) and 7-13 (c) using HEXTE.  Two QPO 
cycles are displayed.  QPOs are detected for orbits 4-6 and 7-13, 
but not for orbits 1-3.  The 95\% confidence upper limit on the 
QPO amplitude for orbits 1-3 is 17.4\%.\label{fig:highenergy}}
\end{figure}

\begin{figure}
\plotone{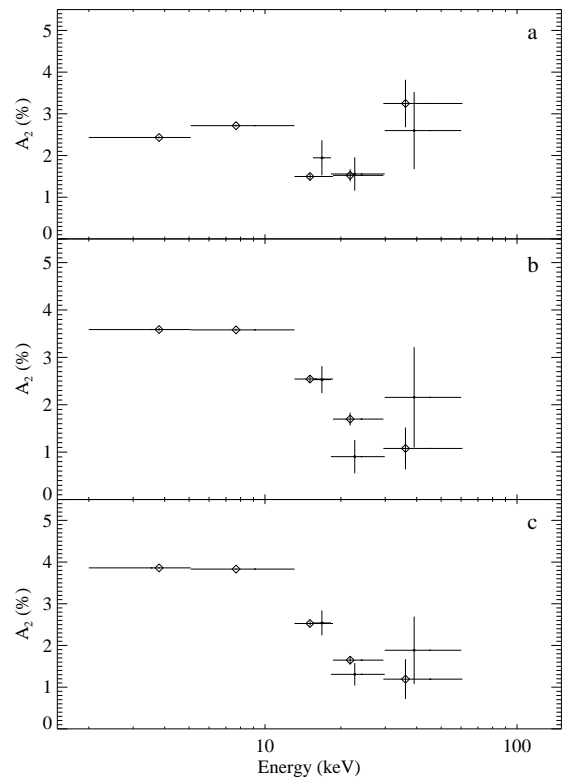}
\vspace{0.5cm}
\caption{The QPO amplitude of the first harmonic ($A_{2}$) vs.~energy
for orbits 1-3 (a), 4-6 (b) and 7-13 (c).  The diamonds mark PCA 
measurements and the points mark HEXTE measurements.\label{fig:amp2}}
\end{figure}

\begin{figure}[t]
\plotone{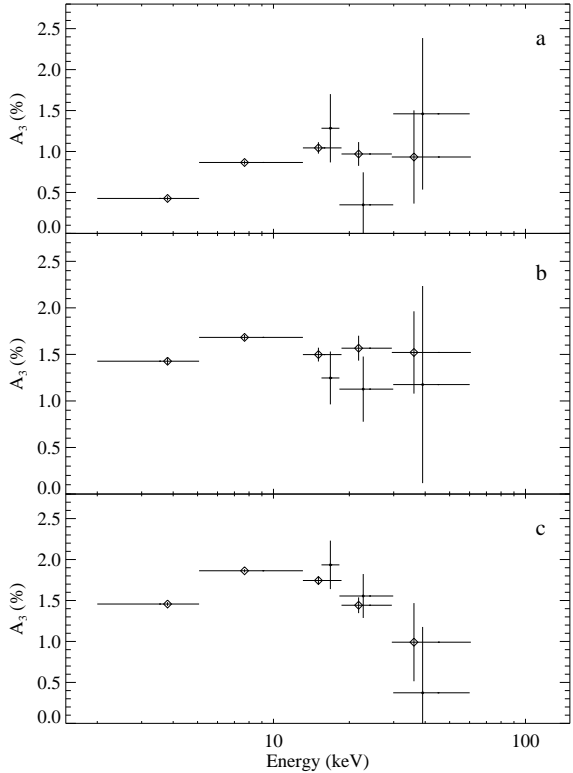}
\vspace{0.5cm}
\caption{The QPO amplitude of the second harmonic ($A_{3}$) vs.~energy
for orbits 1-3 (a), 4-6 (b) and 7-13 (c).  The diamonds mark PCA 
measurements and the points mark HEXTE measurements.\label{fig:amp3}}
\end{figure}

Figure~\ref{fig:amp1} shows the amplitudes of the fundamental
($A_{1}$) for energies from 2 to 124\,keV for orbits 1-3, 4-6
and 7-13 using PCA and HEXTE.  In the region where PCA and HEXTE 
overlap, there is good agreement between the two instruments.  
In all cases, $A_{1}$ increases with energy up to 19\,keV, 
changing from between 11\% and 13\% in the lowest energy band 
up to between 20\% and 24\% in the 13-19\,keV energy band, but
the data do not suggest that $A_{1}$ continues to increase 
above 19\,keV.  While there is strong evidence for a drop 
in $A_{1}$ at higher energies for orbits 1-3, the data do 
not require a decrease in $A_{1}$ above 19\,keV for orbits 
4-6 and 7-13.  However, for orbits 4-6 and 7-13, it should be
noted that $A_{1}$ does not increase with energy at the same 
rate above 19\,keV as it does between 2\,keV and 19\,keV.
Figure~\ref{fig:highenergy} shows the 60-124\,keV folded light 
curves for the three cases.  While the QPO is detected at high 
confidence for orbits 4-6 and 7-13, the QPO is not detected 
for orbits 1-3, and we derive a 95\% confidence upper limit 
($\Delta\chi^{2} = 4$) of 17.4\% on $A_{1}$.  The upper limit 
is found using the model with three harmonic components 
described above and leaving the fit parameters free while 
stepping through values of $a_{1}$.  The results indicate 
that $A_{1}$ is significantly lower for energy bands 
30-60\,keV and 60-124\,keV than for 13-19\,keV and 19-29\,keV.
Figure~\ref{fig:amp2} shows the amplitudes of the first harmonic
($A_{2}$) vs.~energy.  For orbits 4-6 and 7-13, $A_{2}$ is 
between 3.5\% and 4.0\% at low energies and decreases at
higher energies.  For orbits 1-3, $A_{2}$ drops at intermediate 
energies but may increase again above 30\,keV.  
Figure~\ref{fig:amp3} shows the amplitudes of the second harmonic
($A_{3}$) vs.~energy.  The increase in $A_{3}$ at low energies
provides some evidence for similar behavior for the second
harmonic and the fundamental.  In addition, like $A_{1}$, $A_{3}$ 
appears to level off or decrease at higher energies.  For orbits 
7-13, the data strongly suggests that $A_{3}$ peaks around 
10\,keV and decreases at high energies.  Values of $A_{2}$ and 
$A_{3}$ are not well-constrained for the 60-124\,keV energy band 
and are not shown.

\begin{figure}[t]
\plotone{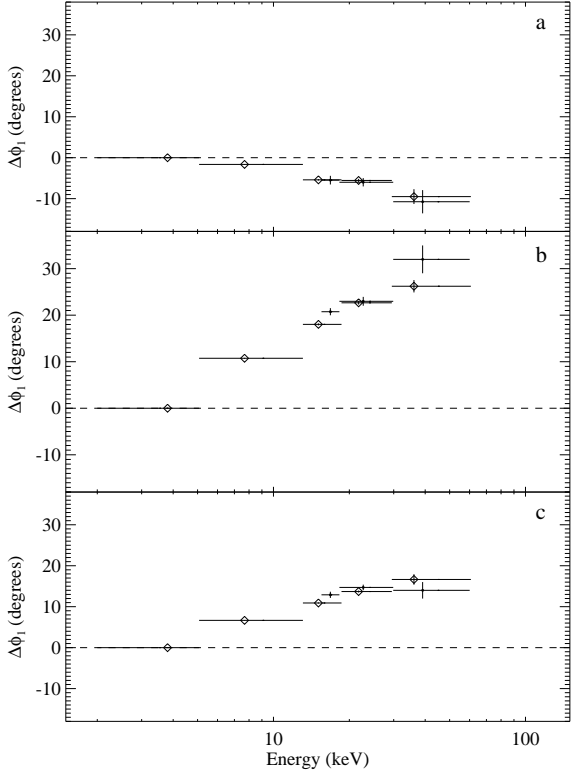}
\vspace{0.5cm}
\caption{The phase lag for the fundamental vs.~energy relative 
to the 2-5\,keV energy band 
(i.e., $\Delta\phi_{1} = \phi_{1}(E) - \phi_{1}$(2-5\,keV))
for orbits 1-3 (a), 4-6 (b) and 7-13 (c).  The diamonds mark PCA 
measurements and the points mark HEXTE measurements.\label{fig:phase1}}
\end{figure}

\begin{figure}[t]
\plotone{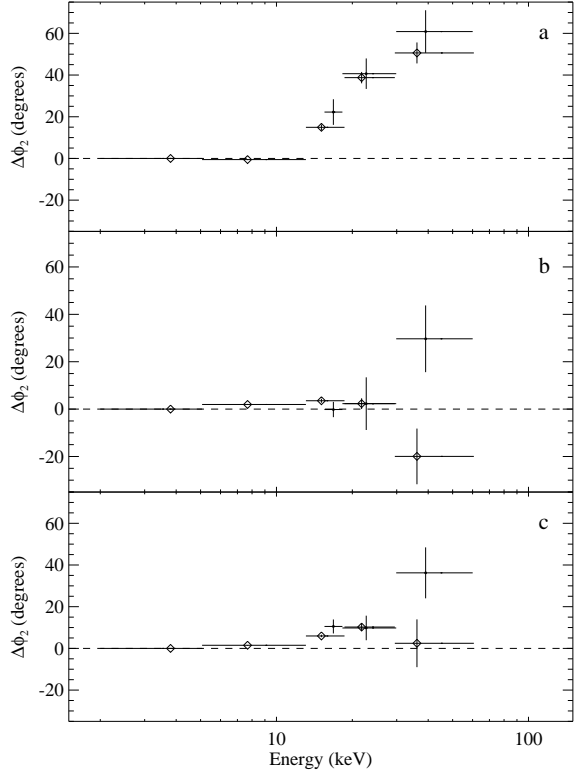}
\vspace{0.5cm}
\caption{The phase lag for the first harmonic vs.~energy relative 
to the 2-5\,keV energy band 
(i.e., $\Delta\phi_{2} = \phi_{2}(E) - \phi_{2}$(2-5\,keV))
for orbits 1-3 (a), 4-6 (b) and 7-13 (c).  The diamonds mark PCA 
measurements and the points mark HEXTE measurements.
\label{fig:phase2}}
\end{figure}

\begin{figure}
\plotone{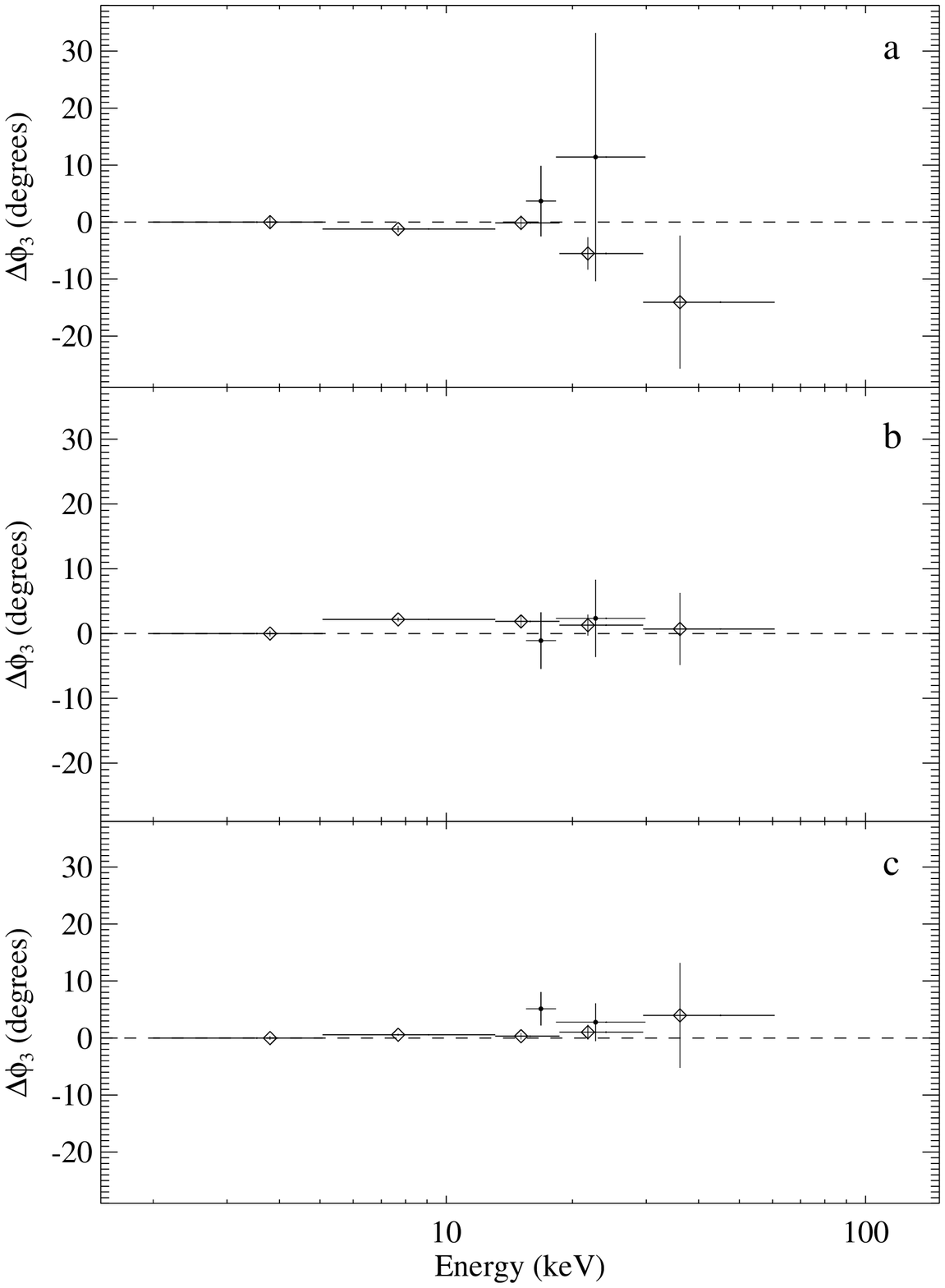}
\vspace{0.5cm}
\caption{The phase lag for the second harmonic vs.~energy relative 
to the 2-5\,keV energy band 
(i.e., $\Delta\phi_{3} = \phi_{3}(E) - \phi_{3}$(2-5\,keV))
for orbits 1-3 (a), 4-6 (b) and 7-13 (c).  The diamonds mark PCA 
measurements and the points mark HEXTE measurements.
\label{fig:phase3}}
\end{figure}

Figure~\ref{fig:phase1} shows the phase lag of the fundamental 
($\Delta\phi_{1}$) for different energies up to 60\,keV relative 
to the 2-5\,keV energy band 
(i.e., $\Delta\phi_{1} = \phi_{1}(E) - \phi_{1}$(2-5\,keV))
for orbits 1-3, 4-6 and 7-13 using PCA and HEXTE.  As we found 
for the amplitude, the PCA and HEXTE measurements are consistent.
The values of $\Delta\phi_{1}$ we find for the 5-13\,keV energy
band are consistent with the phase lag values found by 
Reig et al.~(2000) using cross-spectral techniques.  
For orbits 1-3, our results show that the negative phase lags 
reported by Reig et al.~(2000) for the 5-13\,keV energy band 
become more negative at higher energies.  For orbits 4-6 and 
7-13, the phase lags are positive and increase steadily with
energy.  Figure~\ref{fig:phase2} shows the phase lag of the
first harmonic ($\Delta\phi_{2}$) relative to the 2-5\,keV
energy band.  For orbits 1-3, positive phase lags are seen
for the first harmonic above 13\,keV, in contrast to the
negative phase lags found for the fundamental.  For orbits 4-6 
and 7-13, the phase lags for the first harmonic are also 
positive, but they are significantly smaller than for orbits 
1-3 and somewhat smaller than the phase lags found for the 
fundamental for orbits 4-6 and 7-13.  Figure~\ref{fig:phase3} 
shows the phase lag of the second harmonic ($\Delta\phi_{3}$) 
relative to the 2-5\,keV energy band.  There is some evidence 
that the second harmonic has negative lags for orbits 1-3; 
however, the significance of the detection is marginal.  For 
orbits 4-6 and 7-13, there is marginal evidence for positive 
lags.  Note that the phase lag for the second harmonic is not 
well constrained by HEXTE for the 30-60\,keV energy band and 
is not shown.

\section{Discussion}

Our results for GRS~1915+105 focus on the energy dependence
of the QPO amplitude and phase lags for the QPO fundamental 
and its first two harmonics.  In this section, we discuss how our 
results relate to some of the proposed models to explain the X-ray 
timing properties of BHCs.  Recent work by 
Lehr, Wagoner \& Wilms (2000) provides one of the only detailed 
predictions for the energy dependence of the QPO amplitude for 
BHCs.  Assuming that the hard X-rays are produced by 
Comptonization in a corona with a radial temperature gradient, 
Lehr et al.~(2000) predict that, in some cases, the QPO amplitude 
will decrease at high energies, similar to what we observe for
GRS~1915+105.  In their Monte Carlo simulations, 
Lehr et al.~(2000) assume that the QPO is due to the flux 
modulation of soft seed photons which are subsequently Compton 
upscattered in a corona.  The total X-ray flux emitted is 
the combination of the flux from regions where the QPO is active 
and the flux from other regions where the soft photons are 
unmodulated.  In a given energy band, the predicted QPO 
amplitude is given by the ratio of the flux that emerges in
this energy band from the region where the QPO is active to the 
total flux in this energy band.  If the corona has a temperature 
gradient, and is hottest in the region where the QPO is active, then 
a QPO amplitude that increases with energy is predicted because 
the highest energy photons come from the region where the QPO 
is active.  However, if the QPO is active in a region where the 
corona is cooler, then a drop in the QPO amplitude at high energies 
is predicted.  Within this model, observing the amplitude as a 
function of photon energy can be used to gain insight into where 
the QPO originates.  If it is assumed that the electron temperature 
decreases with distance from the compact object (e.g., if the energy 
deposition in the corona becomes less efficient at larger radii), then 
QPOs originating in the accretion disk at a greater distance from the 
compact object will exhibit a larger drop in amplitude at high energies.

Results for the GRS~1915+105 QPOs appear to be qualitatively
consistent with the simulations of Lehr et al.~(2000) if it is
assumed that lower frequency QPOs originate further out in
the accretion disk, which would be the case if the QPO frequency
is related to the dynamical time scale where the QPO is produced.
For GRS~1915+105, Cui (1999) finds that the 0.067\,Hz QPO
amplitude peaks between 7\,keV and 10\,keV.  Our results show 
that the amplitude of the 2.6-3.0\,Hz QPO (orbits 1-3) peaks 
between 13\,keV and 30\,keV.  The fact that the amplitude of 
this QPO peaks at a higher energy than the amplitude of the 
0.067\,Hz QPO is consistent with a QPO production site closer 
to the compact object.  Finally, the 67\,Hz QPO observed for
GRS~1915+105 is also consistent with this picture since
its amplitude has only been observed to increase with energy 
in the PCA energy range where it is detected 
(Morgan et al.~1997).  However, the highest energy band where 
the 67\,Hz QPO has been detected is only 15-25\,keV.

It is difficult to explain the negative phase lags observed 
for the QPO fundamental for orbits 1-3.  If hard X-rays 
are produced by inverse Comptonization, then one expects
the hard X-rays to be delayed relative to the soft X-rays, 
which results in positive phase lags as is observed for
orbits 4-6 and 7-13.  Reig et al.~(2000) find that the
phase lags for the continuum noise are also negative in 
some cases when the QPO phase lags are negative, and any 
model to explain the negative QPO phase lags should also
be able to explain the negative continuum phase lags.
Such a model, motivated by observations of GRS~1915+105,
has been proposed by Nobili et al.~(2000).  Their model 
assumes that the optical depth of the Comptonizing region 
increases as the inner edge of the disk moves closer to 
the compact object so that Compton down-scattering (instead 
of up-scattering) dominates when the inner edge of the disk 
is close to the compact object.  At this point, the model
of Nobili et al.~(2000) provides a possible explanation 
for the behavior of the continuum, but more work is 
necessary to see if the model can reproduce all the observed 
QPO properties.

Although the model of Nobili et al.~(2000) may be able
to explain the negative phase lags for the QPO fundamental, 
it is not clear if the model will be able to explain the
simultaneous positive phase lags for the first harmonic.
This phenomenon (alternating negative and positive phase
lags between the QPO fundamental and its first harmonic)
has also been observed for the BHC XTE~J1550--564
(\cite{whv99};~\cite{sobczak2000}).  An explanation
for this behavior has been suggested by 
B$\ddot{\rm o}$ttcher \& Liang (2000).  They assume that 
inverse Comptonization is still the dominant process, but, 
in addition, they assume that the light travel time between 
the location where the soft photons are emitted and the 
Comptonization region is longer than half of the fundamental 
QPO period, making a hard lag appear to be a soft lag.  
Depending on the distance between the source of soft 
photons (i.e., the optically thick accretion disk) and the 
Comptonization region, the model can produce positive 
lags (i.e., apparent hard lags) or negative lags (i.e., apparent 
soft lags).  Negative lags are predicted for higher frequencies, 
which is consistent with the properties of the low-frequency 
QPOs in GRS~1915+105.  In certain cases, the model of 
B$\ddot{\rm o}$ttcher \& Liang (2000) predicts negative 
lags for the fundamental, positive lags for the first harmonic 
and negative lags for the second harmonic.  For 
GRS~1915+105, we observe positive lags for the first 
harmonic, and, although our results suggest that the phase 
lags for the second harmonic may be negative, the evidence 
for this is marginal.  While this model may provide 
explanations for the observed QPO phase lags, it is not 
clear if it can explain the negative phase lags observed 
for the continuum noise.

\section{Summary and Conclusions}

Our study of the broadband properties of the low-frequency 
QPOs in GRS~1915+105 provides several new results.  We 
developed a QPO folding technique allowing us to study
the detailed properties of the GRS~1915+105 QPOs.  We
successfully tested the folding technique using simulations, 
and, where comparison is possible, our results are 
consistent with previous results that relied on standard
Fourier techniques.  We report the first detection of
QPOs with HEXTE, making it possible to measure QPO
amplitudes and phase lags up to 124\,keV.  The highest 
energy band where we achieved a detection is 60-124\,keV.  
In the energy range where there is overlap between the PCA 
and HEXTE, we find good agreement between the two instruments 
for the QPO properties.  While our results confirm that 
the QPO amplitude increases with energy up to 19\,keV, they 
also indicate that the QPO amplitude does not increase at 
the same rate above 19\,keV as it does at lower energies.  
For orbits 1-3, the QPO amplitudes in the 30-60\,keV and 
60-124\,keV energy bands are significantly less than in the 
13-19\,keV and 19-29\,keV energy bands.  Also, for orbits 1-3, 
we confirm the negative phase lag for the QPO fundamental 
reported by Reig et al.~(2000), and we report a positive 
phase lag for the first harmonic.  Variation in the phase 
difference between sinusoidal components leads to changes 
in the shape of the QPO profile with energy and over time.  
Our results show that, in some cases, the QPO profile remains 
unchanged for more than 3\,d even as the QPO frequency varies 
by 17\% (orbits 7-13).  Also, the QPO profile can change 
significantly on time scales less than 5 or 6\,d.  Although the 
physical models developed by Lehr et al.~(2000), 
Nobili et al.~(2000) and B$\ddot{\rm o}$ttcher \& Liang (2000)
and discussed here each can explain aspects of the 
GRS~1915+105 timing properties, it is not clear if any of 
the models can reproduce all the observed timing properties.  
More detailed comparisons between the data and the model 
predictions are necessary.

\acknowledgements

JAT would like to acknowledge participants of the 
Rossi 2000 meeting at Goddard Space Flight Center in 
2000 March for useful discussions and J$\ddot{\rm o}$rn 
Wilms for comments on a draft of this paper.

\clearpage

\begin{deluxetable}{lccccc}
\footnotesize
\tablecaption{GRS 1915+105 Observations from October 1997 \label{tab:obs}}
\tablewidth{0pt}
\tablehead{\colhead{Orbit} & \colhead{MJD\tablenotemark{a}} 
& \colhead{Exposure(s)\tablenotemark{b}} & \colhead{QPO Frequency (Hz)} 
& \colhead{PCA Rate (cps)\tablenotemark{c}} & \colhead{HEXTE Rate (cps)\tablenotemark{d}}}
\startdata
1 & 50729.347 & 3392 & 2.940 & 9349 & 62\nl
2 & 50729.407 & 1599 & 2.627 & 8986 & 61\nl
3 & 50730.414 & 3408 & 2.730 & 8815 & 60\nl
4 & 50735.558 & 1960 & 0.844 & 7342 & 80\nl
5 & 50737.421 & 2943 & 1.013 & 7338 & 75\nl
6 & 50737.488 & 2158 & 1.064 & 7308 & 74\nl
7 & 50743.301 & 1119 & 1.398 & 7319 & 70\nl
8 & 50743.335 & 3161 & 1.358 & 7299 & 72\nl
9 & 50743.416 & 2745 & 1.394 & 7262 & 70\nl
10 & 50743.486 & 2377 & 1.437 & 7323 & 70\nl
11 & 50746.553 & 2103 & 1.397 & 7334 & 71\nl
12 & 50746.622 & 1975 & 1.481 & 7336 & 69\nl
13 & 50746.692 & 1639 & 1.609 & 7432 & 68\nl
\tablenotetext{a}{Modified Julian Date at the midpoint of the unocculted part of the orbit.}
\tablenotetext{b}{On-source time for the PCA.}
\tablenotetext{c}{2-30\,keV count rate for 5 PCUs after background subtraction.}
\tablenotetext{d}{15-124\,keV HEXTE count rate after deadtime correction and 
background subtraction.}
\enddata
\end{deluxetable}

\begin{deluxetable}{lccc}
\footnotesize
\tablecaption{Phase Differences Between Sinusoidal 
Components\tablenotemark{a} \label{tab:phases}}
\tablewidth{0pt}
\tablehead{\colhead{Orbits} & \colhead{$\phi_{1}-\phi_{2}$} 
& \colhead{$\phi_{1}-\phi_{3}$} & \colhead{$\phi_{2}-\phi_{3}$}}
\startdata
1-3 & $256.27\pm 0.32$ & $256.75\pm 1.10$ & $0.48\pm 1.13$\nl
4-6 & $226.49\pm 0.32$ & $232.58\pm 0.49$ & $6.08\pm 0.53$\nl
7-13 & $239.00\pm 0.20$ & $242.23\pm 0.32$ & $3.23\pm 0.35$\nl
\tablenotetext{a}{The phase differences are for the 2-5\,keV
	folded light curve and are given in degrees.}
\enddata
\end{deluxetable}

\end{document}